\documentclass[sigconf, screen]{acmart}
\usepackage{tabularx}
\usepackage{pifont}
\usepackage{xspace}
\usepackage{enumitem}

\newcommand{\xmark}{\ding{55}}
\newcommand{\Fig}[1]{Fig.~\ref{#1}}

\newcommand{\Tab}[1]{Tab.~\ref{#1}}
 
\newcommand{\Sec}[1]{Sec.~\ref{#1}}
\newcommand{\GTSim}{GTSim\xspace}
\newcommand{\GTSimTitle}{GPU-Tile-Sim\xspace}
\newcommand{\GTSimFull}{\GTSimTitle{} (\GTSim)}
\definecolor{revisionblue}{RGB}{0,0,0}
\newcommand{\Rev}[1]{}
\newenvironment{revision}{}{}

\makeatletter
\renewcommand\footnotetextcopyrightpermission[1]{}
\makeatother
\AtBeginDocument{%
  }

\setcopyright{none} 
\copyrightyear{2026}
\acmYear{2026}
\acmDOI{XXXXXXX.XXXXXXX}

\acmConference[MICRO 2026]{The 58th IEEE/ACM International Symposium on Microarchitecture}{October 31--November 04, 2026}{Athens, Greece}
\acmISBN{978-X-XXXX-XXXX-X/XX/XX}

\begin{document}

\title{\GTSimTitle: A Tile-Centric GPU Simulation Framework for LLM Hardware-Software Co-Design}
% \subtitle{\normalsize{MICRO 2026 Submission
%     \textbf{\#1260} -- Confidential Draft -- Do NOT Distribute!!}}

% GPU Tile Sim
%\author{\normalsize{MICRO 2026 Submission
 %   \textbf{\#NaN} -- Confidential Draft -- Do NOT Distribute!!}}

\author{Yitong Ding}
\affiliation{%
  \institution{Shanghai Jiao Tong University}
  \city{}
  \country{China}}
\email{dingyitong@sjtu.edu.cn}

\author{Jiawei Huang}
\affiliation{%
  \institution{Shanghai Jiao Tong University}
  \city{}
  \country{China}}
\email{huangjiawei@sjtu.edu.cn}

\author{Renyang Guan}
\affiliation{%
  \institution{Shanghai Jiao Tong University}
  \city{}
  \country{China}}
\email{guanrenyang@alumni.sjtu.edu.cn}

\author{Yangjie Zhou}
\authornote{Corresponding authors.}
\affiliation{%
  \institution{National University of Singapore}
  \city{}
  \country{Singapore}}
\email{yj_zhou@nus.edu.sg}

\author{Zihan Liu}
\authornotemark[1]
\affiliation{%
  \institution{Shanghai Jiao Tong University}
  \city{}
  \country{China}}
\email{altair.liu@sjtu.edu.cn}

\author{Yu Feng}
\affiliation{%
  \institution{Shanghai Jiao Tong University}
  \city{}
  \country{China}}
\email{y-feng@sjtu.edu.cn}

\author{Shixuan Sun}
\affiliation{%
  \institution{Shanghai Jiao Tong University}
  \city{}
  \country{China}}
\email{sunshixuan@sjtu.edu.cn}

\author{Mingyi Guo}
\affiliation{%
  \institution{Shanghai Jiao Tong University}
  \city{}
  \country{China}}
\email{guo-my@cs.sjtu.edu.cn}

\author{Jingwen Leng}
\affiliation{%
  \institution{Shanghai Jiao Tong University}
  \city{}
  \country{China}}
\email{leng-jw@cs.sjtu.edu.cn}

\author{Jian Weng}
\affiliation{%
  \institution{King Abdullah University of Science and Technology (KAUST)}
  \city{}
  \country{Saudi Arabia}}
\email{jian.weng@kaust.edu.sa}

\renewcommand{\shortauthors}{Yitong Ding et al.}


%%%%%% -- PAPER CONTENT STARTS-- %%%%%%%%
\begin{abstract}
%Modern LLM hardware-software co-design increasingly relies on highly optimized GPU kernels whose performance is determined by fine-grained dependencies and overlap between computation and data movement. This places new demands on GPU performance models: instruction-driven simulators are costly to adapt to rapidly evolving architectures, while analytical models are often too coarse to capture these kernels.

Modern LLM (large language model) workloads increasingly rely on  optimized GPU kernels through hardware-software co-design.
These kernels achieve high-performance through fine-grained dependency scheduling and computation-memory overlap. 
As such, they incur new challenges on existing GPU performance models. 
Instruction-driven simulators are costly to adapt to evolving architectures, while analytical models are too coarse to capture kernels' characteristics.
We propose \GTSimFull, a tile-centric GPU simulation framework for LLM hardware-software co-design. The key insight is that modern LLM kernel performance is governed less by individual instruction latency than by the dependency structure that controls execution order and overlap. Accordingly, \GTSim represents kernel execution as a warp-level tile graph whose nodes capture tile-level operations and whose edges encode data and ordering constraints. 
Using this representation, we design an automatic tile-graph frontend and a graph-driven simulation backend.
We evaluate \GTSim on representative GEMM, attention, and end-to-end LLM inference workloads.
On A100 and H100 across both conventional and highly optimized kernels, \GTSim achieves high performance-modeling accuracy (MAPE, Mean Absolute Percentage Error, 1.22\%--8.71\%).
%By explicitly modeling the dependency structure that governs execution order and overlap in modern kernels, 
 We further extend \GTSim to Blackwell with preliminary validation, and demonstrate its effectiveness in analyzing software and architectural design choices.

\end{abstract}


%\begin{CCSXML}
%<ccs2012>
%  <concept_id>00000000.0000000.0000000</concept_id>
%  <concept_desc>Do Not Use This Code, Generate the Correct Terms for Your Paper</concept_desc>
%  <concept_significance>500</concept_significance>
%  %<concept_id>00000000.00000000.00000000</concept_id>
%  <concept_desc>Do Not Use This Code, Generate the Correct Terms for Your Paper</concept_desc>
%  <concept_significance>300</concept_significance>
%  %<concept_id>00000000.00000000.00000000</concept_id>
%  <concept_desc>Do Not Use This Code, Generate the Correct Terms for Your Paper</concept_desc>
%  <concept_significance>100</concept_significance>
 % <concept_id>00000000.00000000.00000000</concept_id>
%  <concept_desc>Do Not Use This Code, Generate the Correct Terms for Your Paper</concept_desc>
%  <concept_significance>100</concept_significance>
%</ccs2012>
%\end{CCSXML}

%\ccsdesc[500]{Do Not Use This Code~Generate the Correct Terms for Your Paper}
%\ccsdesc[300]{Do Not Use This Code~Generate the Correct Terms for Your Paper}
%\ccsdesc{Do Not Use This Code~Generate the Correct Terms for Your Paper}
%\ccsdesc[100]{Do Not Use This Code~Generate the Correct Terms for Your Paper}


\maketitle

\section{Introduction}

% co-design important
% 做协同设计能更好利用性能 硬件啥特性 -> 软件设计 展开讲tensor cores，TMA等等
Modern GPUs are key platforms for LLM workload
execution. To better exploit their peak performance, 
%Hardware-software co-design has become increasingly important for high-performance LLM workloads on modern GPUs. Recent GPU generations continue to raise peak compute throughput. To help software approach this theoretical performance more closely,
recent GPUs introduce features enabled by hardware-software co-design, such as Tensor Memory Accelerator, distributed share memory, and Tensor Memory~\cite{nvidia_ampere_whitepaper_pdf,nvidia_hopper_whitepaper_pdf,nvidia_blackwell_gtc25}. As Tensor Core throughput continues to increase and execution becomes coarser-grained, sustaining high utilization requires kernels to keep compute pipelines fed through tiling, kernel fusion, and multi-stage software pipelines that overlap data movement with matrix computation. At the same time, increasingly asynchronous data-movement and on-chip sharing mechanisms encourage programming styles based on producer--consumer coordination, warp specialization, and explicit data placement~\cite{cutlass,dao2023flashattention2,flashattention3,flashattention4,deepseek_v3,luo2025clusterfusion}. As a result, performance increasingly depends on co-designing software execution with these evolving hardware mechanisms.

%先讲performance model有哪些
To guide such hardware-software co-design, GPU performance models must predict how software and architectural choices interact to affect performance. Existing approaches fall into two broad categories: instruction-driven simulators and analytical models, with the latter further divided into interval-based models and mapping-based frameworks. These approaches embody a key tradeoff between fidelity and extensibility. Instruction-driven simulators~\cite{bakhoda2009gpgpusim,khairy2020accelsim,binkert2011gem5,power2015gem5gpu} execute generation specific instruction-level program traces through detailed microarchitectural models, which gives them high fidelity but also makes them costly to extend to new architectures. Interval-based analytical models~\cite{wang2020mdm,cha2022gcom,cao2025amali} also rely on instruction-level traces, but replace explicit instruction simulation with higher-level interval abstractions. This improves efficiency, yet makes them less suitable for mainstream recent-GPU execution paradigms, where techniques such as warp specialization violate their core representative-warp assumption. Mapping-based frameworks~\cite{parashar2019timeloop,kwon2019maestro,zheng2023tileflowfusion,zhang2024llmcompass} describe programs through loop- or mapping-level abstractions, making them easier to retarget across architectures but often too coarse to represent fine-grained asynchronous execution and warp-specialized coordination explicitly. Overall, existing approaches are either too tightly coupled to low-level instruction semantics or too coarse to capture the explicit dependency of modern GPU kernels, making it difficult to support both new execution paradigms and rapid hardware evolution.

% graph general: 适配： 软件technique 和 硬件 feature和硬件粒度
% 例子：TMA 天然tile。 DSM 通过连边。讲清楚这样做有什么优势，不仅是为什么这样做
In this paper, we present \GTSimFull, a tile-centric GPU simulation framework for LLM hardware-software co-design. The core observation is that modern GPU kernel performance is governed more by dependencies across tile-level computation and data movement than by individual instruction latency. To address the limitations of prior approaches, \GTSim models kernel execution as a dependency-driven warp-centric tile graph, in which nodes represent tile operations executed by warps or cooperating warp groups, and edges explicitly encode dataflow and ordering constraints. This abstraction makes modern kernel execution explicit \textbf{without tying the model to generation-specific instruction streams, yet still capturing fused kernels, software pipelines, and warp specialization.}

%讲一些细节，定义加一些细节，模拟也加一些。讲一些道理出来（怎么解决挑战）
Based on this representation, \GTSim first uses an automatic frontend to extract tile graphs from a tile-based domain-specific language (DSL), currently instantiated with TileLang IR~\cite{wang2025tilelang}. It then uses a graph-driven backend that schedules ready nodes in the tile graph directly rather than advancing instruction streams. The backend estimates the latency of each node in the tile graph using throughput-oriented compute, memory, and NoC abstractions. By driving execution from explicit node dependencies while abstracting hardware at the throughput level rather than simulating instruction pipelines, \GTSim remains lightweight yet accurately captures scheduling, contention, and overlap. To capture concurrency across warps, nodes can be decomposed into sub-operations at a granularity aligned with pipeline width, and a configurable scheduler selects which ready sub-operations to issue onto the corresponding modeled resources.

% 直接分开来讲：kernel level -> e2e
% case study : 先说做case study，再说extend to B200
% open source
We evaluate \GTSim on representative GEMM, attention, and end-to-end LLM inference workloads on A100 and H100, covering both conventional and highly optimized kernels~\cite{flashattention3,flashdecoding,deepseek_v3}. \GTSim achieves MAPE (Mean Absolute Percentage Error~\cite{hyndman2006forecast_accuracy}) from 1.22\% to 6.50\% and consistently outperforms prior analytical models such as TileFlow~\cite{zheng2023tileflowfusion} and LLMCompass~\cite{zhang2024llmcompass} on both conventional and optimized kernels. 
For the Llama-3-8B inference workload, we achieve MAPE from 7.06\% to 8.71\%. Based on an H100-like architecture, we use \GTSim to study software pipelining and NoC-enabled fusion with topology-aware mapping. We further adapt \GTSim to Blackwell with preliminary validation, and use it to explore Blackwell architecture-aware kernel design.

%自动构建单独讲
%最后两点可以合并
Our contributions are as follows:
\begin{itemize}[leftmargin=*]
\item We present \GTSim, a GPU simulation framework that models GPU kernel execution as a dependency-driven warp-centric tile graph, explicitly capturing computation, data movement, execution groups, and dependencies in a unified representation.
\item We develop a tile-graph frontend that supports automatic extraction from a tile-based DSL and a graph-driven simulation backend that executes the graph using throughput-oriented compute, memory, and NoC abstractions, avoiding full instruction-level simulation while retaining extensibility to various architectures.
\item We validate \GTSim on GEMM, attention, and end-to-end LLM inference workloads on A100 and H100, showing low error and better accuracy than prior analytical models, and demonstrate its utility for hardware-software co-design through case studies and preliminary B200 validation.
\end{itemize}

\section{Background and Motivation}
\label{sec:background-and-motivation}

In this section, we describe LLM workloads and their impact on GPU architectures and programming techniques.
We then present the resulting challenges for GPU performance modeling.
%\textcolor{red}{This section reviews the workload, architectural, and programming trends behind modern GPU kernels, and the resulting challenges for performance modeling.}

\begin{table*}[!t]
\centering
\small
\setlength{\tabcolsep}{4pt}
\caption{Architecture-related features across recent NVIDIA GPU generations.}
\vspace*{-0.2cm}
\label{tab:tensor_core_evolution}
\begin{tabular}{lcccc}
\toprule
Feature & Pre-Ampere & Ampere & Hopper & Blackwell \\
\midrule
Tensor Core operation granularity
& Warp-scoped
& Warp-scoped
& Warpgroup-level (\texttt{wgmma})
& Warpgroup / CTA-pair \\

Key low-precision support (examples)
& FP16 (mixed)
& FP16 / BF16
& FP16 / BF16 / FP8
& FP16 / BF16 / FP8 and lower \\

Async global$\rightarrow$shared mechanism
& N/A
& \texttt{cp.async}
& TMA
& TMA \\

Inter-SM cooperation
& N/A
& N/A
& DSMEM
& DSMEM \\

Operand storage for Tensor Core
& Registers
& Registers
& SMEM for \texttt{wgmma}
& Tensor Memory (TMEM) + SMEM \\
\bottomrule
\end{tabular}
\end{table*}

\subsection{LLM Workload}

\begin{figure}[t]
    \centering
    \includegraphics[width=0.9\linewidth]{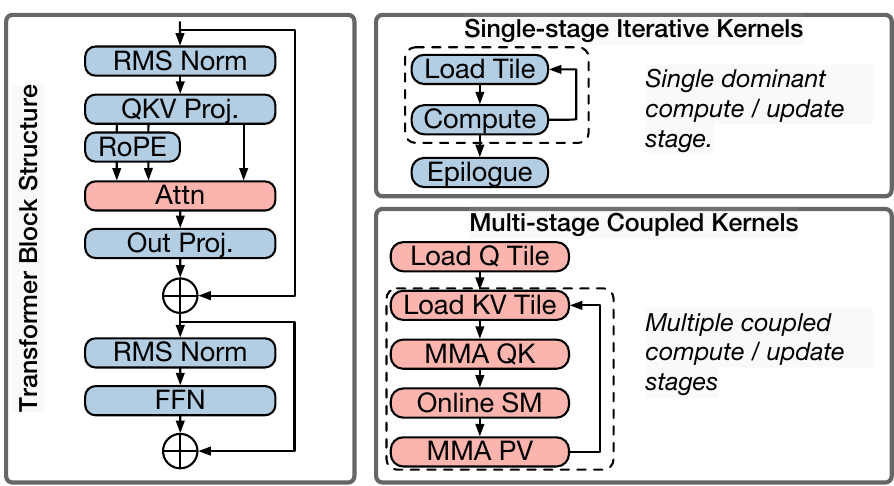}
    \vspace*{-0.3cm}
    \caption{Representative kernel patterns in LLM workloads. Left: a typical transformer block. Right: single-stage iterative and multi-stage coupled kernels.}
    \label{fig:llm-workload-dependency}
\end{figure}

\begin{figure}[t]
    \centering
    \includegraphics[width=0.9\linewidth]{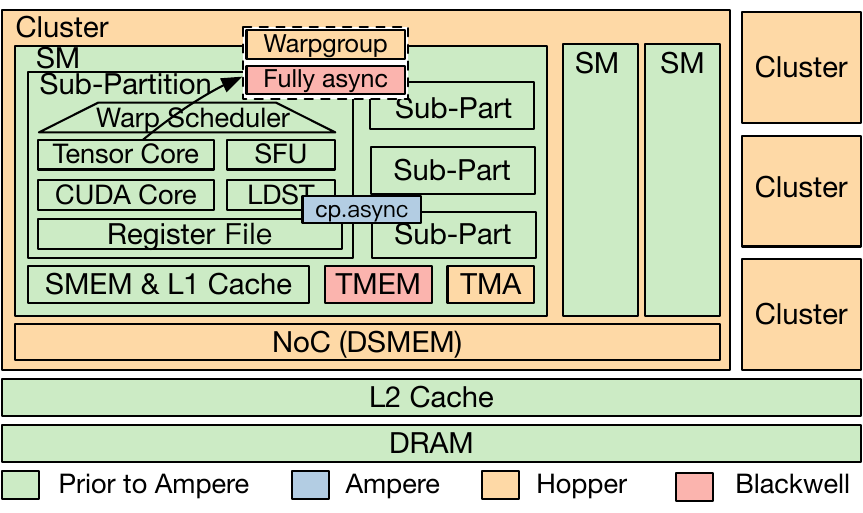}
    \vspace*{-0.3cm}
    \caption{Architectural evolution across GPU generations.}
    \label{fig:hardware-model}
\end{figure}

Although LLM architectures continue to evolve, the performance-critical GPU kernels have not changed fundamentally: they still center on a small number of kernel types~\cite{vaswani2017attention,brown2020gpt3,touvron2023llama}. The important distinction lies in the dependency structure that governs execution order, overlap, and coordination. As shown in \Fig{fig:llm-workload-dependency}, a typical transformer block broadly exhibits two representative kernel patterns. We classify them by the structure of their basic execution unit: one category contains a single dominant compute/update stage, while the other contains multiple coupled stages.

\textbf{Single-stage iterative kernels.} A large class of LLM kernels has a basic execution unit with a single dominant compute/update stage and exhibits relatively regular dependencies, where data movement and computation are repeatedly interleaved with limited cross-stage coordination. Typical examples include projection and feed-forward layers, which are commonly implemented using optimized GEMM kernels, fused variants, and emerging low-precision variants such as FP8 GEMM~\cite{vaswani2017attention, cutlass, fp8_formats}. In these kernels, execution usually follows repeated load-compute iterations with relatively uniform dependencies across stages, allowing relatively consistent overlap between data movement and computation.

\textbf{Multi-stage coupled kernels.} Another important class of LLM kernels has a basic execution unit with multiple coupled stages, which lead to stronger cross-stage dependencies and tighter interaction among data movement, computation, and synchronization. 
Attention-style kernels are such examples: FlashAttention and its subsequent variants fuse multiple stages into a single kernel, which includes online intermediate stages while overlapping memory movement with computation~\cite{dao2022flashattention, dao2023flashattention2, flashattention3, flashattention4}. 
Other variants such as Flash-Decoding and FlashMLA further introduce more irregular data access and reuse patterns~\cite{flashdecoding, deepseek_v3} to improve performance. In these kernels, execution is more strongly constrained by dependencies, making available overlap more limited.

Overall, performance-critical LLM kernels demonstrate a small set of representative execution patterns, with different dependency structures that govern execution order, overlap, and coordination.

\subsection{GPU Architecture Evolution}

Across recent GPU generations, the basic SM-based organization remains the same: kernels execute on streaming multiprocessors (SMs), where computation is organized around warps, thread blocks, on-chip memory, and Tensor Cores~\cite{nvidia_volta_whitepaper,cuda_prog_guide,volkov_latency_hiding}. Each SM has multiple subpartitions, and each subpartition has an independent scheduler that issues instructions for the warps assigned to it. What changes across generations is the execution model itself, especially how computation, data movement, and coordination are organized and overlapped. The key trend is a shift from tightly coupled, warp-centric execution toward increasingly decoupled, cooperative, and explicitly managed multi-stage execution. \Fig{fig:hardware-model} summarizes this evolution and the hardware support behind it.

\textbf{Prior to Ampere} GPUs (e.g., Volta~\cite{nvidia_volta_whitepaper} and Turing~\cite{nvidia_turing_whitepaper}) largely follow a tightly coupled execution model, where computation and data movement are coordinated at warp scope with limited overlap. Concretely, Tensor Core execution is mostly warp-scoped and synchronous, while data movement is orchestrated through explicit load/store instructions and register staging.

\textbf{Ampere} GPU marks the first clear step toward decoupled intra-kernel pipelines by allowing data movement to proceed more independently from computation~\cite{nvidia_ampere_whitepaper_pdf}. This shift is enabled by \texttt{cp.async}, which supports asynchronous global-to-shared transfers and makes staged pipelines with partial copy-compute overlap practical.

\textbf{Hopper} GPU extends this trend by turning decoupling into coarser-grained and more distributed execution across larger execution groups~\cite{nvidia_hopper_whitepaper_pdf}. Specifically, warpgroup-level tensor-core execution (\texttt{wgmma}), bulk asynchronous transfer (TMA), and DSMEM-based cluster cooperation together enable coordinated execution across multiple warps and even across SMs.

\textbf{Blackwell} GPU pushes the execution model further toward explicitly managed multi-stage execution by more aggressively separating computation, data movement, and operand storage~\cite{nvidia_blackwell_gtc25}. In particular, \texttt{tcgen05} and TMEM further decouple operand residency from registers, enabling more asynchronous tensor-core execution, while CTA-pair execution within a tensor processing cluster (TPC) enables larger-scale cooperative tensor-core execution.

Overall, these architectural changes move the GPU execution model from tightly coupled warp-scoped execution toward more decoupled, distributed, and explicitly managed execution.

\subsection{GPU Programming Techniques}
\label{sec:gpu-programming-techniques}

The key trend in modern GPU programming is a shift toward more explicit coordination of data movement, computation, and synchronization to create effective overlap. This trend is driven jointly by the demands of modern LLM workloads and by recent architectural support for asynchronous data movement and decoupled execution resources (e.g., \texttt{cp.async}, TMA)~\cite{nvidia_ampere_whitepaper_pdf, nvidia_hopper_whitepaper_pdf}. These trends are reflected in several important programming techniques.

\textbf{Kernel fusion} has become a central technique to reduce memory traffic and improve locality in LLM workloads. Representative examples include FlashAttention~\cite{dao2022flashattention, dao2023flashattention2, flashattention3, flashattention4}, FlashDecoding~\cite{flashdecoding}, and FlashMLA~\cite{deepseek_v3}, which fuse multiple operators (e.g., attention score computation, softmax, and matrix multiplication) into a single kernel. By keeping intermediate results in on-chip memory (shared memory) and eliminating redundant global memory accesses, these approaches organize execution into tightly coupled stages with explicit data dependencies across fused operations.

\textbf{Software pipelining} is widely used to overlap data movement and computation within a kernel. Leveraging asynchronous copy mechanisms, modern kernels employ multi-stage pipelines (e.g., double or multi-buffering in shared memory) to prefetch data for future computation stages while current stages execute~\cite{dao2022flashattention, dao2023flashattention2}. This creates a steady-state execution pattern where data transfer and compute proceed concurrently, and performance depends on maintaining balanced pipeline stages and avoiding stalls.

\textbf{Warp specialization} further enhances this overlap by assigning different warps or warp groups to distinct roles, such as data movement, tensor core computation, or synchronization~\cite{nvidia_hopper_whitepaper_pdf, nvidia_blackwell_gtc25}. This technique decouples producer and consumer roles at the warp level, improves parallelism, and mitigates pipeline stalls~\cite{chen2025tawa}. 
%It can also reduce register pressure and enable more aggressive tiling, improving data reuse and reducing redundant memory traffic. As a result, execution is no longer dominated by independent warps, but by coordinated producer--consumer relationships among specialized warps.

Overall, modern GPU programming techniques increasingly rely on explicit coordination among data movement, computation, and synchronization. As a result, kernel performance is determined not only by individual operations, but also by how effectively different stages are overlapped and coordinated.

\subsection{GPU Performance Modeling Challenges}

Since performance-critical LLM kernels increasingly rely on explicit coordination of data movement, computation, and synchronization, a useful GPU performance model must therefore capture these dependency-driven execution patterns while remaining extensible to new hardware features.
%The discussion above highlights two key requirements for modern GPU performance modelling. First, performance-critical LLM kernels differ mainly in the dependency structure that governs execution order, overlap, and coordination. Second, recent architectures and programming techniques increasingly rely on more explicit coordination of data movement, computation, and synchronization, together with rapidly evolving execution mechanisms. A useful performance model must therefore capture these dependency-driven execution patterns while remaining extensible to new hardware features.}

However, existing GPU performance models do not satisfy these requirements simultaneously. 
There are broadly two types of such models: instruction-driven simulators and analytical models. 
In-struction-driven simulators (e.g., GPGPU-Sim~\cite{bakhoda2009gpgpusim}, Accel-Sim~\cite{khairy2020accelsim}, gem5 and its GPU extensions~\cite{binkert2011gem5,power2015gem5gpu}) take instruction streams such as PTX or SASS traces as input and drive execution by simulating the behavior of each instruction through detailed microarchitectural components such as warp scheduling, pipelines, and memory systems, thereby achieving high fidelity. 
Analytical models can be further divided into interval-based models and mapping-based frameworks.
Interval-based models (e.g., MDM~\cite{wang2020mdm}, GCoM~\cite{cha2022gcom}, AMALI~\cite{cao2025amali}) similarly rely on PTX or SASS traces, but approximate execution by partitioning a representative warp's execution into stall-delimited intervals and estimating cycles interval by interval. They then use that representative behavior to estimate overall execution, instead of explicitly simulating every instruction. Mapping-based frameworks (e.g., Timeloop~\cite{parashar2019timeloop}, MAESTRO~\cite{kwon2019maestro}, TileFlow~\cite{zheng2023tileflowfusion}, LLMCompass~\cite{zhang2024llmcompass}) model program behavior through loop- or map-ping-level abstractions, making them easier to adapt but less explicit about fine-grained execution.

While these models report accurate performance in certain cases, existing approaches still have limited support for key requirements of modern GPU kernels, including kernel fusion, asynchronous pipelines, warp specialization, and architecture support. We summarize these limitations in \Tab{tab:model_comparison}.

%Rather than comparing models by methodology, we focus on their ability to support key capabilities required by modern GPU kernels. \Tab{tab:model_comparison} summarizes this comparison.

\begin{table}[t]
\centering
\small
\setlength{\tabcolsep}{4pt}
\caption{Comparison of existing GPU performance models.}
\vspace*{-0.2cm}
\label{tab:model_comparison}
\begin{tabular}{l l c c}
\toprule
Model &
\shortstack{Granularity\\\& Paradigm} &
\shortstack{Programming Tech.\\(KF, AP, WS)} &
\shortstack{Arch. Supp.\\(AHB)} \\
\midrule

Accel-Sim~\cite{khairy2020accelsim} &
Inst / Sim &
\checkmark\ \checkmark\ $\triangle$ &
\checkmark\ $\triangle$\ $\triangle$ \\

AMALI~\cite{cao2025amali} &
Inst / Ana &
\checkmark\ \checkmark\ \xmark &
\checkmark\ $\triangle$\ $\triangle$ \\

LLMCompass~\cite{zhang2024llmcompass} &
Kernel / Ana &
\xmark \ $\triangle$\ \xmark &
\checkmark\ $\triangle$\ $\triangle$ \\

TileFlow~\cite{zheng2023tileflowfusion} &
Tile / Ana &
\checkmark\ \xmark \ \xmark &
$\triangle$\ $\triangle$\ $\triangle$ \\

\textbf{Ours} &
Tile / Sim &
\checkmark\ \checkmark\ \checkmark &
\checkmark\ \checkmark\ \checkmark \\

\bottomrule
\end{tabular}

\vspace{2pt}
\footnotesize
KF: Kernel Fusion; AP: Asynchronous Pipeline; WS: Warp Specialization. \\
Inst: Instruction-level; Ana: Analytical; Sim: Simulation. \\
\textcolor{revisionblue}{\checkmark: supported; $\triangle$: approximate support or substantial extension required; \xmark: not supported.} \\
AHB denotes Ampere, Hopper, and Blackwell support levels.
\end{table}

% \begin{table*}[t]
% \centering
% \small
% \setlength{\tabcolsep}{4pt}
% \caption{Capability comparison of existing GPU performance models.}
% \label{tab:model_comparison}
% \begin{tabular}{l l c c c c c}
% \toprule
% Model & Model &
% Kernel Fusion &
% Fine-grained Async &
% Warp-specialized &
% Ampere Feature &
% Hopper/ Blackwell \\
% Name & Type  &
% Support &
% Modeling &
% Pipeline &
% Support &
% Feature Support \\
% \midrule
% Accel-Sim~\cite{khairy2020accelsim} & Instruction-driven & \checkmark & \checkmark & \triangle & \checkmark & \triangle \\
% AMALI~\cite{cao2025amali} & Interval-based & \checkmark & \checkmark & \xmark & \checkmark & \triangle \\
% LLMCompass~\cite{zhang2024llmcompass} & Mapping-based & \xmark & \triangle & \xmark & \checkmark & \triangle \\
% TileFlow~\cite{zheng2023tileflowfusion} & Loop-based & \checkmark & \xmark & \xmark & \triangle & \triangle \\
% \textbf{Ours} & Graph-driven & \checkmark & \checkmark & \checkmark & \checkmark & \checkmark \\
% \bottomrule
% \end{tabular}

% \begin{tablenotes}
% \small
% \item \checkmark: supported; $\triangle$: partially supported or heavy modification needed; \xmark: not supported.
% \end{tablenotes}
% \end{table*}

\textbf{Programming techniques support.}
\begin{revision}
\Rev{D1} Modern GPU kernels increasingly rely on kernel fusion, asynchronous pipelines, and warp specialization, which make performance depend on fine-grained dependencies, overlap, and non-uniform warp or warp-group behavior. Instruction-driven simulators can in principle capture these effects, but supporting such features often requires substantial updates beyond adding new instructions, especially in dependency handling, scheduling, and execution pipelines. For example, Hopper's \texttt{wgmma} introduces warpgroup-level commit-wait style weak ordering, while Blackwell's \texttt{tcgen05} introduces single-thread issue semantics for tensor-core execution~\cite{nvidia_hopper_whitepaper_pdf,nvidia_blackwell_gtc25}. Interval-based models are particularly challenged by warp specialization because they estimate execution from representative warp intervals and therefore struggle when specialized producer and consumer warps exhibit heterogeneous progress and synchronization patterns. Mapping-based frameworks operate at a higher abstraction level through operator mappings, loop transformations, tiling, and resource binding. However, these abstractions generally lack a uniform execution representation for composing arbitrary fused operations and explicitly encoding heterogeneous warp roles, synchronization events, and non-data ordering constraints. They therefore cannot directly represent, or require substantial extensions to approximate, fine-grained asynchronous and warp-specialized execution.
\end{revision}

\textbf{Hardware extensibility.}
Emerging architectures continually introduce new tensor instructions, execution groups, memory spaces, and asynchronous data-movement mechanisms. For instruction-driven simulators and interval-based analytical models, supporting these features requires updating trace generation or parsing and extending the model to new instruction semantics and execution behavior.
This high engineering cost makes academic simulators lag behind hardware generations: the official GPGPU-Sim repo only supports tested up to CUDA 11-era software support~\cite{gpgpusim_repo}, and does not natively support newer Hopper and Blackwell GPUs.
% such as \texttt{wgmma} or \texttt{tcgen05}. 
Mapping-based frameworks remain easier to adapt, but their coarse hardware abstractions usually only approximate new features rather than modeling their performance impact directly.

\textbf{Summary. }
Overall, no existing approach simultaneously captures dependency-driven execution patterns such as fine-grained asynchronous execution and warp-specialized coordination, while also remaining readily extensible to emerging architectural features and maintaining both accuracy and efficiency.

\section{\GTSimTitle{} Overview}

In this section, we present an overview of \GTSimFull, a graph-driven modeling framework designed to capture dependency-driven execution in modern GPU kernels while remaining extensible to evolving hardware mechanisms.

\textbf{Our key insight is that modern GPU kernel performance is governed primarily by how data movement, computation, and synchronization depend on and overlap with one another, rather than by individual instruction latency.} Leveraging this insight, we represent a kernel execution process as a warp-centric \textbf{tile graph}, in which nodes capture warp-level tile operations and edges encode data and ordering dependencies. This representation makes overlap, coordination, and different execution roles explicit within a unified abstraction, rather than recovering them indirectly from instruction streams or coarse mappings.

\begin{figure}[t]
    \centering
    \includegraphics[width=\linewidth]{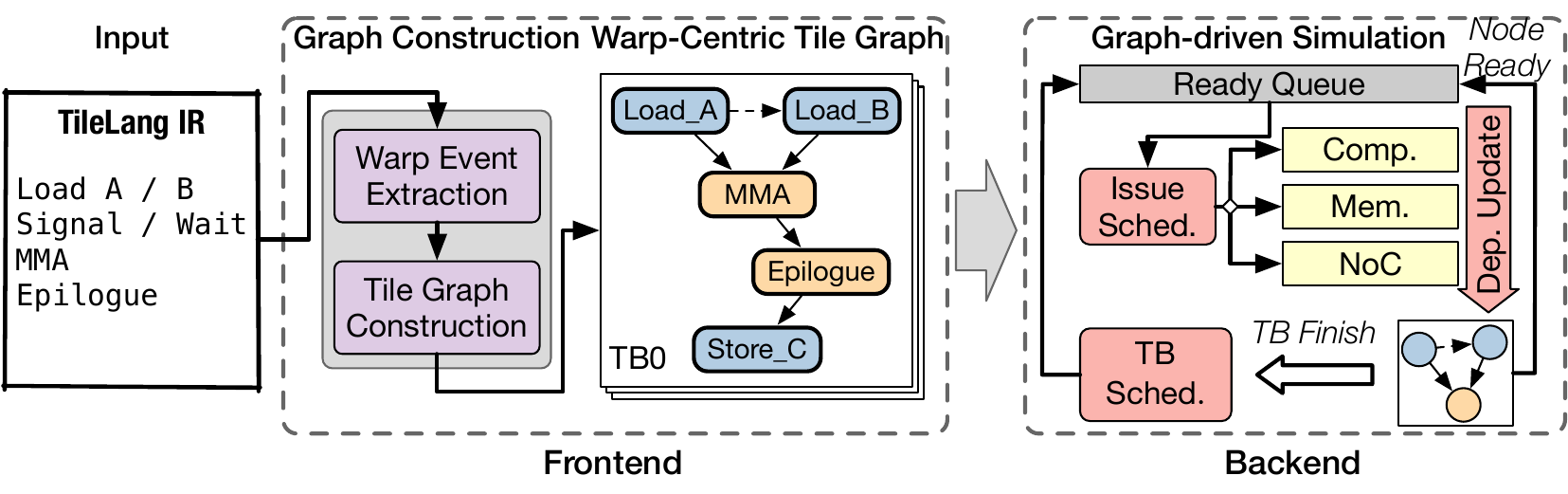}
     %  \vspace*{-0.2cm}
    \caption{Overview of \GTSim.}
    \label{fig:overview}
\end{figure}

\Fig{fig:overview} illustrates the overall workflow of \GTSim. 
The frontend of \GTSim, described in \Sec{sec:tile-graph-frontend}, takes a warp-centric tile graph as the input.
This tile graph captures explicit dependencies and synchronization among warps.
The frontend also supports tile-graph construction from TileLang IR~\cite{wang2025tilelang}, which allows broad coverage of LLM workloads and GPU programming techniques.
 %as input and converts it into  warp-event extraction and  in the frontend . 
 The backend (\Sec{sec:graph-driven-simulation-backend}) takes in the tile graph and performs graph-driven simulation based on user-specified hardware model that consists of compute, memory, NoC, and various warp/threadblock schedulers.
 
% repeatedly selects ready nodes, dispatches them onto throughput-oriented compute, memory, and NoC models, and uses dependency updates to activate successor nodes, while a threadblock scheduler manages block residency and completion.

Together, this workflow enables \GTSim to make dependency structure explicit from frontend input through backend execution, directly model overlap and coordination across heterogeneous execution roles, and remain extensible to emerging GPU mechanisms within a unified framework.

\section{Tile Graph Frontend}
\label{sec:tile-graph-frontend}

%从硬件的角度开始写->引出tile graph

\begin{figure*}[t]
    \centering
    \includegraphics[width=0.95\linewidth]{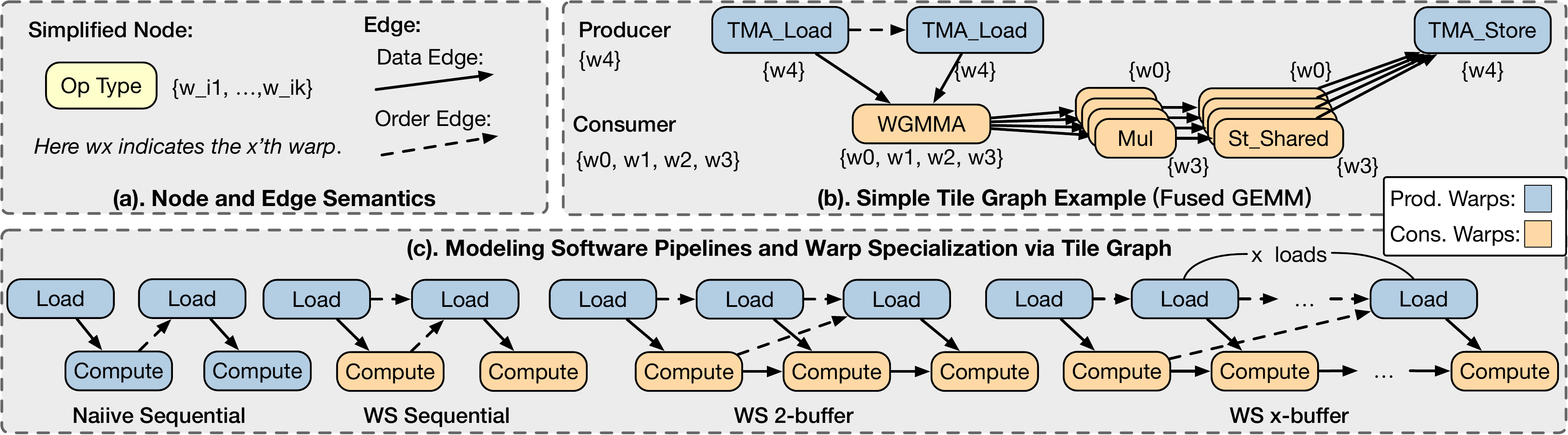}
    \vspace*{-0.3cm}
    \caption{Tile graph examples. (a) Simplified node and edge semantics. (b) A simple fused-GEMM tile graph with producer and consumer warp groups. (c) Software pipeline organizations, including warp-specialized (WS) variants.}
    % expressed through the same tile-graph abstraction.}
    \label{fig:tile-graph-example}
\end{figure*}

In this section, we describe the frontend of \GTSim that adopts the warp-centric tile graph to represent a kernel, which serves as the input to graph-driven simulation backend.

%is responsible for constructing the warp-centric tile graph that . The frontend takes TileLang IR as input and automatically constructs the corresponding tile graph.

\subsection{Tile Graph Abstraction}
\label{sec:tile-graph-abstraction}

As shown in \Sec{sec:background-and-motivation}, modern GPU kernels increasingly rely on explicit dependencies, asynchronous overlap, and coordination across specialized execution roles. 
These features motivate an abstraction that must explicitly capture three aspects: the tensor tiles being manipulated, the warp-level execution groups that perform each operation, and the ordering constraints that coordinate execution.

\textbf{Warp-Centric Tile Graph. } We model GPU execution at \emph{warp granularity}. This choice provides a balance between expressiveness and simulation efficiency. A finer-grained thread-level abstraction would reintroduce much of the complexity of instruction- and address-level modeling, while a coarser threadblock-level abstraction would obscure the heterogeneous roles of different warps and fail to represent warp-specialized pipelines or warp-group collective operations. Since the warp is the basic scheduling unit on modern GPUs, a warp-centric abstraction naturally captures the execution roles and coordination patterns that matter for performance.

Based on this design choice, we represent a kernel as a directed acyclic graph $G = (V, E)$, where each node $v \in V$ indicates a warp-level operation, and each edge $e \in E$ indicates a relationship between operations. To support graph-driven simulation, we augment each node and edge with execution semantics as follows.
%each node is associated with an execution group, an operation descriptor, and a tile descriptor, while each edge is described by an edge descriptor.

\textbf{Node Semantics. } Each node represents a \emph{tile-level execution event} and records which logical warp or group of warps executes it. This information is essential because, in modern kernels, the identity of the executor is part of the execution behavior itself: some operations are carried out by a single warp, some by specialized producer or consumer warps, and others by multiple cooperating warps, such as warp-group tensor-core instructions (e.g., \texttt{wgmma}). This allows the abstraction to represent both warp-specialized execution and collective multi-warp operations.

The \emph{operation descriptor} specifies the semantic type of the operation and the hardware execution unit that serves it, such as CUDA cores, Tensor Cores, SFUs, LD/ST units, or TMA. This allows the backend to map each node directly to the corresponding model without recovering low-level instruction behavior.

The \emph{tile descriptor} specifies the tile produced or accessed by a node, including its storage location, data type, shape, memory layout, and logical coordinates within the parent tensor, which are similar to those used in CUTLASS~\cite{cutlass}.
These information help the backend simulator derives memory transactions from tile-level accesses rather than per-thread addresses.
For example, the tile's coordinates and layout define how a logical tensor region maps to physical memory requests.
Allowing variable tile sizes further enables the abstraction to represent different operational granularities across memory levels, such as large bulk transfers in global or shared memory and fine-grained fragments in registers or tensor-core pipelines.

%The \emph{tile descriptor} describes the tile produced or accessed by the node, including its storage location, data type, shape, layout, and position within the larger tensor. This information is necessary because the backend simulator derives memory transactions from tile-level accesses rather than from per-thread addresses. In particular, the tile position and layout determine how a logical tensor region maps to memory requests, while allowing tile sizes to remain variable enables the same abstraction to represent different granularities of operation across memory levels, such as larger bulk transfers in global/shared memory and smaller fragments in registers or tensor-core pipelines.

\textbf{Edge Semantics. } Edges capture dependencies between nodes. We distinguish two kinds of edges: \emph{data edges} and \emph{order edges}. A \emph{data edge} represents a producer--consumer relationship between two nodes, where one node consumes a tile produced by the other. These edges define the dataflow in the graph and are required for correctness. They explicitly encode dataflow across memory movement, computation, and fused kernels.

An \emph{order edge} represents an ordering constraint that is not fully captured by dataflow. Unlike data edges, order edges do not determine what data is consumed, but constrain when an operation may proceed, often due to synchronization or buffer-management decisions introduced by optimized execution. Such constraints arise in two common cases. First, operations issued by the same warp may need to preserve program order even when they do not exchange data directly. Second, operations executed by different warps may need explicit synchronization through signal/wait pairs, barriers, or buffer-reuse constraints in software pipelines. For example, in a double-buffered pipeline, a producer warp may preload a future tile into a shared-memory buffer slot only after a consumer warp has finished using the previous tile stored in that same slot. These two operations may access different logical tiles and therefore do not form a data edge, yet they must still be ordered to avoid incorrect overlap. Order edges make such constraints explicit.

Overall, the warp-centric tile graph makes tile-level execution groups and dependencies explicit within a unified abstraction, enabling our graph-driven simulation for modeling kernel execution.
% GPU kernels to be represented in a form suitable for .

\subsection{Expressing GPU Programming Techniques}

With these design choices, the tile graph provides a unified representation that captures not only what tiles are produced and consumed, but also which warps execute each operation and how pipelined execution is coordinated. We next show how this abstraction naturally captures key GPU programming techniques in \Sec{sec:gpu-programming-techniques}, including \textbf{kernel fusion}, \textbf{warp specialization}, and \textbf{software pipelining}.

\Fig{fig:tile-graph-example}(a) illustrates a simplified view of node and edge semantics. Each node denotes an operation together with the warp set that executes it, while edges encode either data dependencies (solid arrows) or issue/order constraints (dashed arrows).

\textbf{Kernel Fusion.}
In the tile graph, kernel fusion is represented by eliminating kernel boundaries and expressing cross-stage producer--consumer relationships as direct data edges within one graph. \Fig{fig:tile-graph-example}(b) shows a fused GEMM example integrating TMA loads, warp-group tensor-core computation, epilogue operations, and a final store within a single graph. The producer warp issues \texttt{TMA\_Load} and \texttt{TMA\_Store}, while the consumer warp group executes \texttt{WGMMA}, elementwise multiplication, and shared-memory store operations.

\textbf{Warp Specialization.}
The tile graph represents warp specialization by encoding producer and consumer roles directly in node execution groups, with edges capturing their coordination. In the same example, producer warps perform data movement, while consumer warps perform tensor-core computation and epilogue work.

\textbf{Software Pipelining.}
In the tile graph, software pipelining is represented by staging multiple waves of nodes and using order edges to control buffer reuse without blocking valid overlap. \Fig{fig:tile-graph-example}(c) shows four representative pipeline organizations: Naive Sequential, WS Sequential, WS 2-buffer, and WS $x$-buffer. WS makes the producer--consumer structure explicit, while deeper buffering is expressed by adding load nodes and issue/order edges for future stages. In this way, both overlap and pipeline depth are captured through changes to the same tile-graph abstraction.

\begin{figure}[t]
    \centering
    \includegraphics[width=\linewidth]{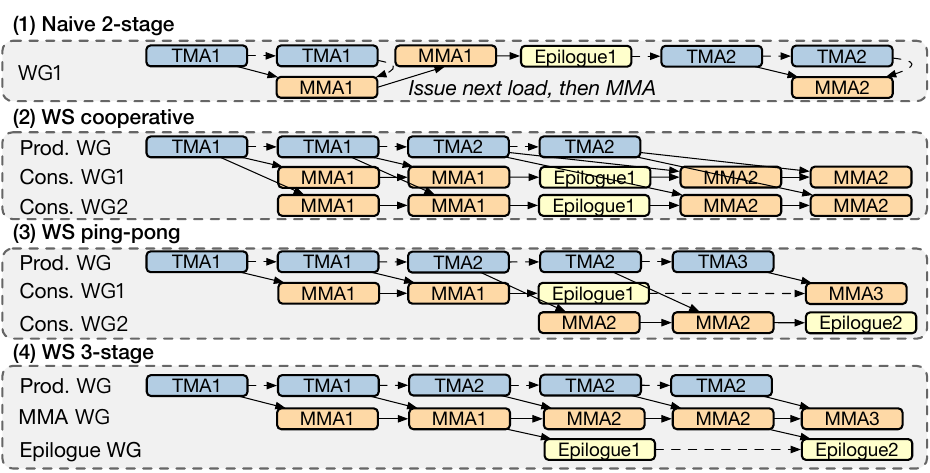}
    \vspace*{-0.3cm}
    \caption{Representative software pipeline organizations for a fused \textbf{GEMM + SiLU} kernel.}
    \label{fig:software-pipeline-setup}
\end{figure}

\Fig{fig:software-pipeline-setup} illustrates four different pipeline organizations of a fused \textbf{GEMM + SiLU} kernel. They differ in execution-group assignment and dependency structure, but they can all be expressed within the same tile-graph abstraction by changing node execution groups and dependency edges rather than the abstraction itself. We quantitatively evaluate these pipeline designs later in \Sec{sec:case-study}.

\subsection{Automatic Tile Graph Generation}
\label{sec:automatic-tile-graph-generation}

\begin{revision}
\Rev{A1} In the currently implemented automatic path, users write kernel source code in the TileLang DSL~\cite{wang2025tilelang}. The TileLang compiler applies software-pipeline and warp-specialization transformations and produces a tile-level IR; the \GTSim frontend then consumes this compiler-generated IR and constructs the tile graph.
\end{revision}
\begin{revision}
\Rev{II} Alternatively, users can directly construct the tile-graph representation defined in \Sec{sec:tile-graph-abstraction} by specifying its node descriptors, execution groups, and data/order edges. Both paths produce the same graph representation consumed by the simulation backend in \Sec{sec:graph-driven-simulation-backend}.
\end{revision}
% and is therefore suitable for \GTSim.

\begin{figure}[t]
    \centering
    \includegraphics[width=1.0\linewidth]{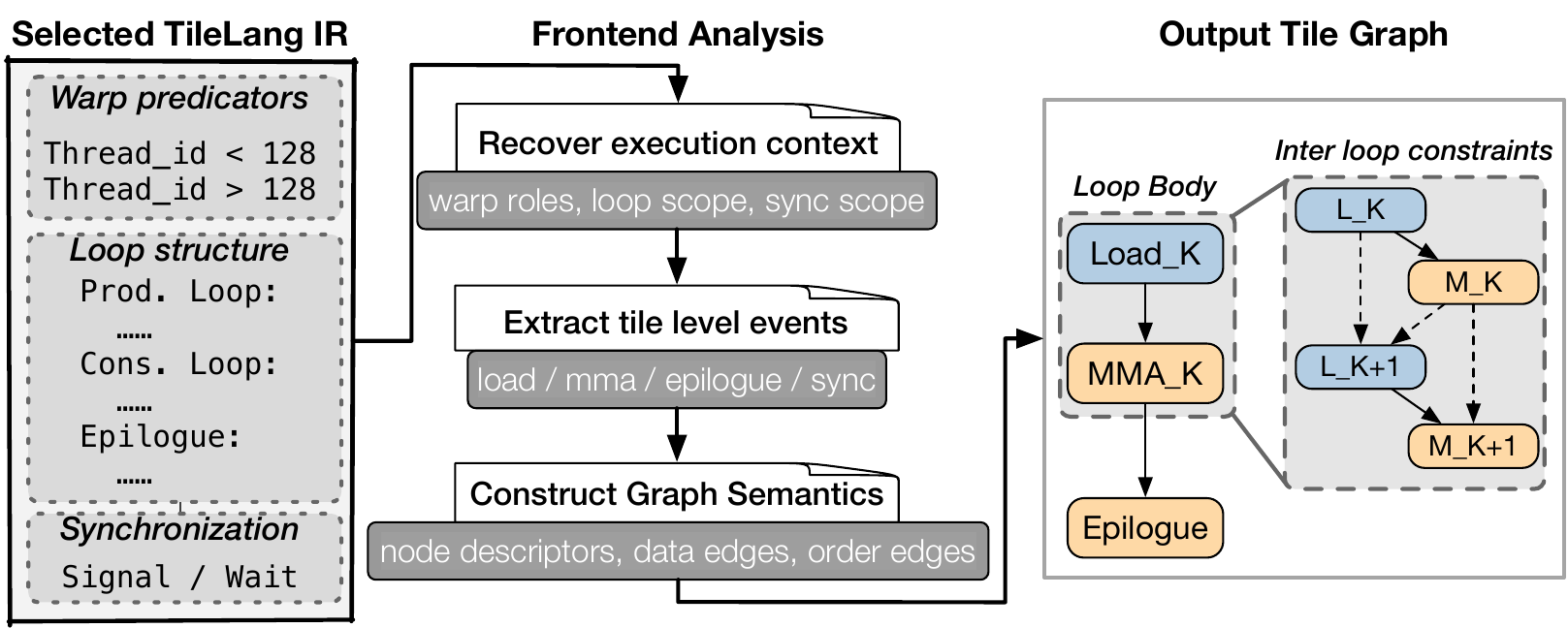}
    \vspace*{-0.3cm}
    \caption{Automatic tile-graph generation from TileLang IR.}
    \label{fig:generation-workflow}
\end{figure}

\Fig{fig:generation-workflow} illustrates the workflow. The \GTSim frontend generates the graph from the compiler-produced tile-level IR rather than low-level CUDA because this IR preserves the execution structure that matters for graph-driven simulation, including warp predicates, loop and pipeline structure, and synchronization. We consume the IR after software-pipeline injection and warp specialization, but before lowering to the instruction level, where these semantics would become less explicit. This choice also improves usability: users can obtain tile graphs directly from existing TileLang source programs without manually reconstructing warp roles, pipeline stages, or synchronization structure.

The frontend analysis in \Fig{fig:generation-workflow} proceeds in three stages. It first performs a structured traversal of the IR to recover execution context from predicates, loop nests, and synchronization constructs, determining warp roles, loop scope, pipeline stages, and cross-warp coordination. It then identifies primitive operations through their IR op kinds, such as tensor-core instructions, TMA loads/stores, scalar or epilogue computations, and explicit synchronization events, and maps each primitive to a tile-graph node. Repeated primitives in unrolled loops are preserved as distinct nodes.

In the final stage, the frontend infers node descriptors and dependencies directly from IR operands and destinations. The operation descriptor is determined from the primitive type and its target execution resource, while the tile descriptor is recovered from the accessed or produced buffer region, including storage location, data type, shape, and logical layout. Data dependencies are built by assigning versioned tile descriptors in a static single assignment (SSA)-like manner~\cite{cytron1991ssa} so that each produced tile has a unique producer, after which consumers are matched to the corresponding producer versions. Order edges are added separately from same-group program order and from explicit synchronization constructs such as signal/wait pairs or barriers, allowing the graph to preserve pipeline ordering and cross-warp coordination without over-constraining independent operations. The resulting graph preserves the fused, warp-specialized, and pipelined execution structure required by graph-driven simulation while remaining much simpler than instruction-level models.

\begin{revision}
\Rev{II} The graph abstraction and backend are language independent, while the frontend path depends on how explicitly the source exposes execution structure. For tile DSLs, a compiler IR can be consumed after pipeline and warp-role lowering, as in our TileLang path. For Triton, TLX demonstrates that role-specialized instruction streams, cross-warp dependencies, and local-memory orchestration can remain explicit in extended TTGIR before downstream lowering~\cite{guan2026tlx}, providing an analogous frontend boundary. For structured CUDA kernels, CuBridge demonstrates a lift from expert CUDA into an executable tile-oriented IR that preserves operations, worker assignments, synchronization, and data/order dependencies~\cite{ma2026cubridge}; such a lifted IR can be translated into the tile graph. When these structures cannot be reliably recovered, users can directly construct the graph by specifying the same fields at a substantially coarser granularity than an instruction trace.
\end{revision}

\section{Graph-Driven Simulation Backend}
\label{sec:graph-driven-simulation-backend}

In this section, we first describe the execution flow of graph-driven simulation backend.
We then detail its throughput-oriented hardware modeling and dependency-driven runtime.
%based on user-specified hardware model that consists of compute, memory, NoC, and various warp/threadblock schedulers.

 %describes how it supports graph-based execution on modern GPUs.

% node如何贴近硬件执行（大node多次issue） 根据硬件约束将node拆开来做。
% 先讲overview 不要和instruction对比，如何去模拟tile graph
% 再讲每一个部件
% 最后讲和instruction的对比以及可扩展性 可扩展性太多了，缩短一点

\begin{figure*}[t]
    \centering
    \includegraphics[width=0.9\linewidth]{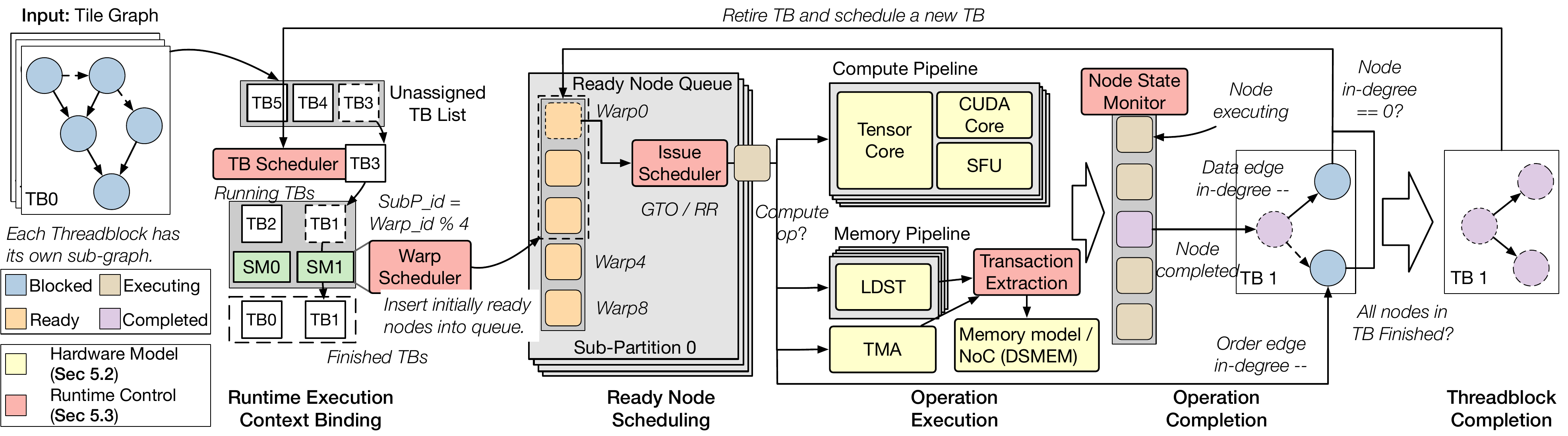}
    \caption{Simulation workflow of the graph-driven simulator.}
    \vspace*{-0.3cm}
    \label{fig:simulation-workflow}
\end{figure*}

\subsection{Backend Execution Flow}

Given a tile graph representation of a GPU kernel, simulation proceeds by executing the graph directly. As illustrated in \Fig{fig:simulation-workflow}, the TB scheduler first binds threadblocks to SMs and places initially ready nodes into per-subpartition ready queues according to their execution groups. Each node then transitions through a lifecycle of \textbf{blocked}, \textbf{ready}, \textbf{issued}, \textbf{executing}, and \textbf{completed}. The issue scheduler repeatedly selects ready nodes from these queues and issues them when the required compute or memory resource is available. At issue time, the simulator releases the outgoing order edges of the node and dispatches it to the corresponding compute or memory pipeline, where execution is tracked until completion.

When a node completes, the simulator releases its outgoing data edges and updates the unresolved-dependency counters of successor nodes. Once a successor's counter reaches zero, it is inserted into the ready queue of the corresponding subpartition and becomes eligible for later issue. In other words, node readiness is determined explicitly from the tile graph: a node becomes ready only when all incoming data and order constraints have been satisfied. This process repeats until all nodes in a threadblock (TB) have completed. The TB is then retired, its SM resources are released, and the TB scheduler can assign a new TB to the freed SM.

This execution model naturally matches the tile-graph abstraction introduced in \Sec{sec:tile-graph-frontend}. Each ready node corresponds to a warp-level tile operation with explicit execution-group, operation, and tile semantics, so the simulator can dispatch it directly to the corresponding hardware model without recovering low-level instruction behavior. The backend therefore focuses on the dominant performance effects of the hardware, while the tile graph supplies the dataflow and ordering structure that governs execution.

\subsection{Throughput-Oriented Hardware Modeling}
\label{sec:throughput-oriented-hardware-modeling}

Once a node is issued, it is serviced by throughput-oriented hardware models, where each resource is characterized by the latency, issue width, and sustained throughput with which it serves tile-level operations. This design aligns naturally with the tile graph because each node already represents a tile-level operation with explicit execution semantics. 
As shown in \Fig{fig:simulation-workflow}, we use this model for hardware execution-resource components including the compute pipeline and the memory pipeline, as well as memory-side service models such as the NoC, which are marked by light-yellow boxes.
% including the  channels that allow different SMs exchange data directly.

\textbf{Compute.}
Compute nodes are mapped to throughput-oriented pipelines corresponding to major execution units such as Tensor Cores, CUDA cores, and SFUs. Each pipeline captures operation latency, pipeline width, and sustained throughput, allowing the simulator to model contention and utilization at the level relevant to tile-graph execution. This level of abstraction is sufficient because each compute node already captures the operation semantics and work granularity relevant to performance modeling.

\textbf{Memory.}
We derive memory accesses from tile descriptors rather than per-thread addresses.
For conventional load/store operations, we expand each tile access into a set of transactions based on tile shape, logical coordinates, memory layout, and the target transaction granularity.
In contrast, we model TMA operations as bulk tile transfers handled by a dedicated TMA front-end, which continuously injects transactions for the entire tile instead of issuing independent load/store requests.
We parameterize the TMA model using microbenchmarks, which show that TMA transfers generate 128B transactions to L2 and sustain an aggregate issue rate of about 100 B/cycle.
We model the L2 cache using a simple fully associative design with an LRU replacement policy, and we determine cache hits and misses based on tile-derived memory accesses mapped to cache lines.
Despite its simplicity, this cache model achieves sufficient accuracy for our evaluation, as it captures most tile-level reuse patterns.
We model L2 and DRAM using latency--bandwidth models that capture service latency, bandwidth limits, and queueing effects, which dominate performance in LLM kernels.

%\textbf{Memory.}
%Memory accesses are derived from tile descriptors rather than per-thread addresses. For conventional load/store operations, a tile access is expanded into a set of transactions according to tile shape, position, layout, and the target transaction granularity. In contrast, TMA operations are modeled as bulk tile transfers handled by a dedicated TMA front-end, which continuously injects transactions for the entire tile rather than treating each transaction as an independent load/store operation. We further parameterize the TMA model using a simple microbenchmark, which shows that TMA transfers generate 128B transactions to L2 and sustain an aggregate issue rate of about 100 B/cycle. The L2 cache and DRAM are both modeled using latency--bandwidth abstractions that capture service latency, bandwidth limits, and queueing effects, which dominate performance for pipelined GPU workloads.

\textcolor{revisionblue}{\textbf{DSMEM / On-chip NoC.}}
\begin{revision}
\Rev{D2} Starting with Hopper, threadblock clusters and distributed shared memory (DSMEM) enable small-scale communication among a group of SMs within one GPU. We model this inter-SM communication as remote SRAM accesses routed through a simplified on-chip SM-cluster NoC with bandwidth, latency, and queueing constraints. Remote accesses carry target SM identifiers in the tile graph, while the NoC model resolves routing and contention. Since \GTSim targets single-GPU kernel simulation, it does not model inter-GPU network communication.
\end{revision}

Together, these resource models capture the dominant compute, memory, and communication effects in tile-graph execution.

\subsection{Dependency-Driven Runtime Control}
\label{sec:scheduling-and-execution}

At runtime, graph execution is advanced by dependency-driven control, which includes readiness tracking, scheduling, issue, state updates, and dependency updates. In \Fig{fig:simulation-workflow}, this part corresponds to the red control components, such as the threadblock, warp, and issue schedulers together with the runtime state-update logic.

Ready nodes are organized into per-subpartition queues and scheduled using standard policies such as greedy-then-oldest~\cite{rogers2012ccws} or round-robin. Scheduling decisions are based on node readiness and resource availability: when a ready node reaches the head of the queue and the corresponding compute or memory resource can accept new work, the node is issued.

Once issued, nodes are dispatched to either compute or memory pipelines. To expose concurrency between warps within throughput-oriented models, a compute node is not treated as a single monolithic request, but instead decomposed into a number of sub-operations determined by the amount of tile work and the sustained throughput of the target pipeline. These sub-operations are injected subject to pipeline width constraints, allowing different nodes to overlap in the same modeled pipeline and improving concurrency while preserving throughput limits. Multi-warp operations (e.g., warp-group tensor core operations) are handled by synchronizing the scheduling entries of all participating warps so that the collective operation is issued consistently across subpartitions.

For a memory node, its issue event triggers transaction generation from the associated tile descriptor. Conventional load/store nodes are decomposed into transactions according to the specified transaction granularity, and each transaction corresponds to a memory-side sub-operation that is scheduled and serviced individually. TMA nodes, in contrast, are issued to the TMA engine as bulk tile transfers, after which 128B transactions are streamed into the memory hierarchy at the calibrated TMA issue rate. These transactions are then serviced by shared memory, L2, DRAM, or DSMEM/NoC models as appropriate. A node is considered complete only after all of its sub-operations or memory transactions have finished. At that point, the simulator releases the outgoing data edges of the completed node and updates the corresponding successor counters. Together with the order-edge updates performed at issue time, this mechanism continuously activates newly ready nodes and drives graph execution forward.

A threadblock is considered complete when all nodes in its tile graph have finished execution, after which the resources allocated to that threadblock on the SM are released, allowing additional threadblocks to be scheduled.

Together, these runtime mechanisms coordinate scheduling, execution, and dependency updates throughout graph execution.

\subsection{Extensibility Advantages}

Compared to instruction-driven simulators, our graph-driven approach is easier to adapt to emerging GPU architectures because execution semantics are represented directly by operation-level nodes and explicit data/order edges. New hardware features can therefore be incorporated by adding new node types, execution resources, or edge rules, without modifying low-level instruction decode, scoreboard, or pipeline logic. For example, supporting a new tensor-core primitive mainly amounts to introducing a new node type and binding it to the appropriate resource model, with only localized updates when new ordering or completion semantics must be expressed. This modularity makes the framework easier to extend as GPU architectures continue to evolve.

\section{Evaluation}

We evaluate \GTSim's accuracy at both the kernel and end-to-end inference level, and compare it against existing analytical models.

\subsection{Experimental Setup}

\textbf{Hardware Configurations. } We evaluate \GTSim across three GPU generations: A100, H100 SXM, and B200. The first two serves as our main evaluation platform while the B200 is mainly used in the  Blackwell case study with a preliminary validation there (\Sec{sec:case-study}). \Tab{tab:eval-config} summarizes the main hardware parameters used in our evaluation. These parameters are derived from official architecture documents, prior benchmarking studies, and our own measurements from targeted microbenchmarks on real devices~\cite{nvidia_ampere_whitepaper_pdf,nvidia_hopper_whitepaper_pdf,luo2025hopper_microbenchmark,jarmusch2025blackwell_microbenchmark}. For real-hardware validation, all A100 and H100 SXM experiments are conducted on systems running CUDA 12.4 and Ubuntu 20.04.6 LTS.

Unless otherwise noted, the validation experiments use full-chip A100 and H100 configurations. For case studies, we use customized setups to highlight the effect of interest, including a single-SM Hopper-like configuration for software pipelining and a NoC-enabled Hopper-like configuration for cluster-based fusion.

\begin{figure*}[t]
    \centering
    \includegraphics[width=0.95\textwidth]{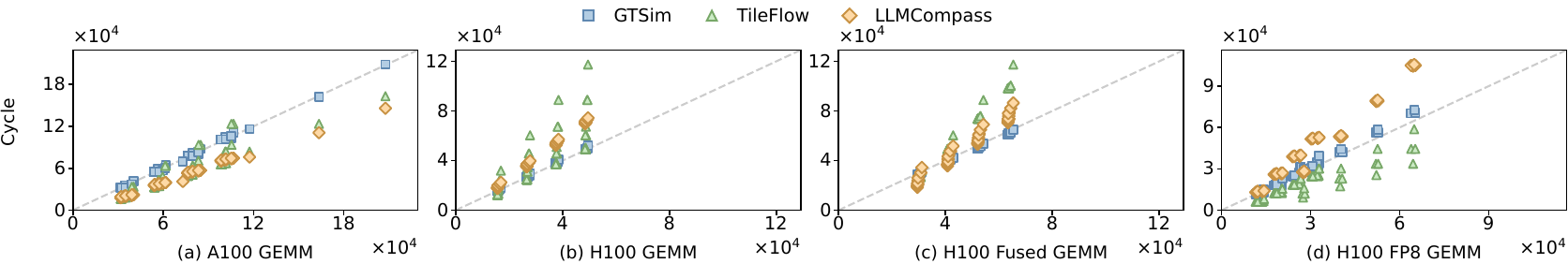}
    \vspace*{-0.3cm}
    \caption{Parity plots for GEMM workloads on A100 and H100, comparing GTSim, TileFlow, and LLMCompass.}
    \label{fig:gemm-validation}
\end{figure*}

\begin{figure*}[t]
    \centering
    \includegraphics[width=0.95\textwidth]{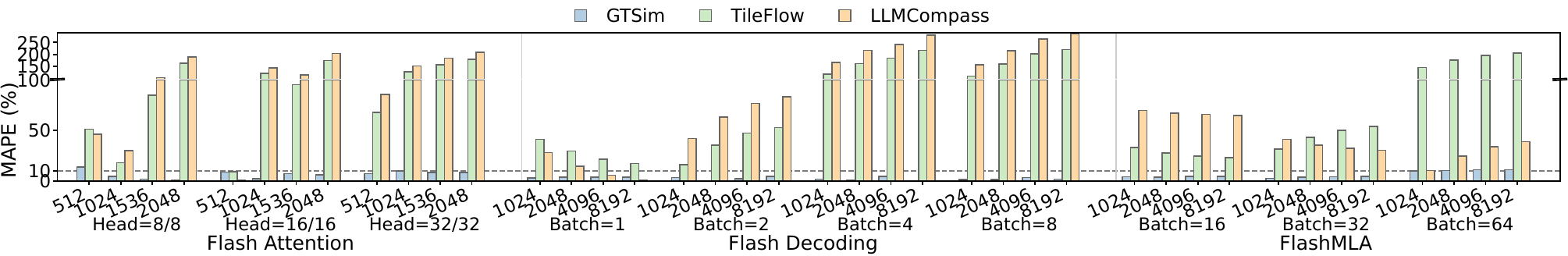}
    \vspace*{-0.3cm}
    \caption{MAPE of GTSim, TileFlow, and LLMCompass on FlashAttention-3, Flash-Decoding, and FlashMLA.}
    \label{fig:attention-validation}
\end{figure*}

\begin{figure}[t]
    \centering
    \includegraphics[width=\linewidth]{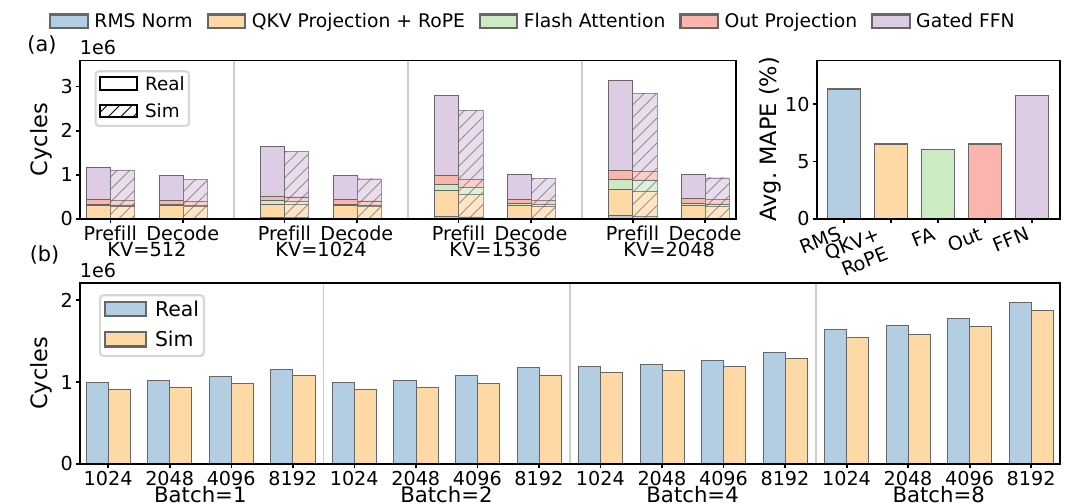}
    \vspace*{-0.5cm}
    \caption{Measured and simulated results of Llama-3-8B inference on H100. (a) Cycle breakdowns for prefill and decode. (b) Decode cycle counts across batch sizes and lengths.}
    \label{fig:llama-3-8b-e2e}
\end{figure}

\begin{table}[t]
\centering
\footnotesize
\setlength{\tabcolsep}{3pt}
\renewcommand{\arraystretch}{1.08}
\caption{GPU configurations used in evaluation.}
\vspace*{-0.3cm}
\label{tab:eval-config}
\begin{tabular}{lccc}
\toprule
\textbf{Parameter} & \textbf{A100} & \textbf{H100 SXM} & \textbf{B200} \\
\midrule
SMs & 108 & 132 & 148 \\
Core / memory freq. (MHz) & 1410 / 1215 & 1980 / 2619 & 1965 / 2000 \\
CUDA cores / SFUs & 64 / 16 & 128 / 16 & 128 / 16 \\
TC throughput (FMAs/clk) & 1024 & 2048 & 4096 \\
SRAM BW / latency (B/cycle / cyc) & 128 / 33 & 128 / 33 & 128 / 33 \\
L2 BW / latency (B/cycle / cyc) & 2300 / 300 & 4300 / 380 & 6000 / 380 \\
DRAM BW / latency (GB/s / cyc) & 1407 / 250 & 3100 / 270 & 8000 / 270 \\
TMA issue rate (B/cycle) & -- & 100 & 100 \\
TMEM BW / latency (TB/s / cyc) & -- & -- & 12 / 28 \\
\bottomrule
\end{tabular}
\end{table}

\textbf{Applications. } We evaluate \GTSim on representative workloads introduced in \Sec{sec:background-and-motivation}, which are GEMM kernels, attention workloads, and their mixed-precision execution. The GEMM workloads include conventional GEMM and fused \textbf{GEMM + SiLU}. The attention workloads include FlashAttention-3~\cite{flashattention3}, Flash-Decoding~\cite{flashdecoding}, and FlashMLA~\cite{deepseek_v3}. To cover mixed-precision behavior, we additionally evaluate FP8 (E5M2) GEMM with FP32 accumulation. Most experiments are conducted on H100, while one GEMM experiment is performed on A100. The real-hardware kernel implementations are primarily based on TileLang~\cite{tilelang_repo}.

\textbf{Compared Models. } We compare \GTSim against two representative mapping-based analytical models, TileFlow~\cite{zheng2023tileflowfusion} and LLMCompass~\cite{zhang2024llmcompass}. TileFlow supports modeling fused kernels through dataflow descriptions, while LLMCompass uses a relatively detailed GPU cost model but only supports single-operator kernels. For the comparison, we adapt TileFlow to match the target algorithm dataflow as closely as possible, and for LLMCompass, we model each fused kernel as a serial execution of its constituent operators. Since both models natively target A100, we further adjust their system parameters to match H100, including the bandwidth of compute units and the memory system. \textcolor{revisionblue}{\Rev{I} We additionally compare runtime and accuracy against the instruction-driven Accel-Sim~\cite{khairy2020accelsim} on A100, the latest architecture supported by the evaluated release.}

\subsection{Kernel-level Evaluation}

Our validation focuses on whether \GTSim can accurately capture how software optimizations change dependency structure, execution order, and overlap in modern GPU kernels, without fully simulating instruction execution.

\textbf{GEMM Kernels. } \Fig{fig:gemm-validation} shows parity plots for four GEMM workloads: A100 GEMM, H100 GEMM, H100 fused GEMM, and H100 FP8 GEMM. Each panel covers 64 $(M, N, K)$ configurations and plots predicted cycles against measured cycles. \GTSim matches real hardware closely across all four settings, achieving MAPE (Mean Absolute Percentage Error~\cite{hyndman2006forecast_accuracy}) values of 2.49\%, 1.22\%, 3.60\%, and 5.40\%, respectively, with Pearson correlation above 0.995 in all cases. In contrast, TileFlow exhibits substantially larger errors, with MAPE ranging from 24.32\% to 34.17\%, while LLMCompass reaches 34.07\% and 36.99\% on A100 and H100 GEMM, and remains less accurate on fused and FP8 GEMM (14.99\% and 40.93\%). The gap becomes especially visible on H100 fused GEMM, whose epilogue is fixed to SiLU, and on H100 FP8 GEMM, which uses E4M3 inputs with FP32 accumulation. This is because \GTSim explicitly models the dependency and overlap structure among TMA transfers, tensor-core execution, and fused epilogue work, while TileFlow and LLMCompass capture these fine-grained execution effects less directly.

\textbf{Attention Kernels. } \Fig{fig:attention-validation} reports the MAPE of \GTSim, TileFlow, and LLMCompass on three representative attention workloads: FlashAttention-3, Flash-Decoding, and FlashMLA. All attention workloads are evaluated on H100. FlashAttention-3 varies the head count and sequence length, Flash-Decoding varies the batch size and KV length with fixed heads$_q$/heads$_{kv}$ = 32/8 and seq$_q$ = 128, and FlashMLA varies the batch size and sequence length with 128 heads. Across these three workloads, \GTSim consistently achieves the lowest error, with average MAPE values of 6.50\%, 2.37\%, and 6.09\%, respectively. In particular, \GTSim remains highly stable on Flash-Decoding, where the worst-case MAPE is only 4.28\%. By contrast, TileFlow and LLMCompass exhibit much larger errors, often exceeding 100\% on FlashAttention-3 and Flash-Decoding, and remain substantially less accurate on FlashMLA. This is because \GTSim captures the full fused execution structure of modern attention kernels, including the dependencies and overlap among tensor-core computation, data movement, and intermediate softmax-related stages. By contrast, TileFlow can represent fused execution flow but uses a coarse cost model that mainly accounts for tensor-core computation and data movement, while LLMCompass only supports single-operator kernels and therefore approximates these fused kernels as a serial execution of separate operators.

\subsection{End-to-end LLM Inference}

\Fig{fig:llama-3-8b-e2e} evaluates \GTSim on end-to-end Llama-3-8B inference on H100. \Fig{fig:llama-3-8b-e2e}(a) compares measured and simulated cycle breakdowns of major decoder-layer kernels under both prefill and decode. Here the batch size is fixed to 1, decode uses $q$ length 128, and the KV length is swept from 512 to 2048. \GTSim tracks both the total execution cost and the per-kernel composition well across all settings, with an overall average MAPE of 8.71\%. The right subpanel of \Fig{fig:llama-3-8b-e2e}(a) further shows that kernel-level average MAPE stays within 11.3\% for all major components, indicating that the simulator remains accurate not only at the end-to-end level but also for the major kernels that dominate decoder execution.

\Fig{fig:llama-3-8b-e2e}(b) further focuses on decode-only inference, using batch sizes from 1 to 8 and KV lengths from 1024 to 8192 with fixed $q$ length 128. Across these 16 configurations, \GTSim achieves an overall average MAPE of 7.06\%, with a worst-case error of 8.87\%. The error remains stable as both batch size and KV length increase, showing that \GTSim can robustly model long-KV decode behavior beyond individual kernels.

\subsection{\textcolor{revisionblue}{Analysis and Discussion}}

\begin{figure}[t]
    \centering
    \includegraphics[width=0.95\linewidth]{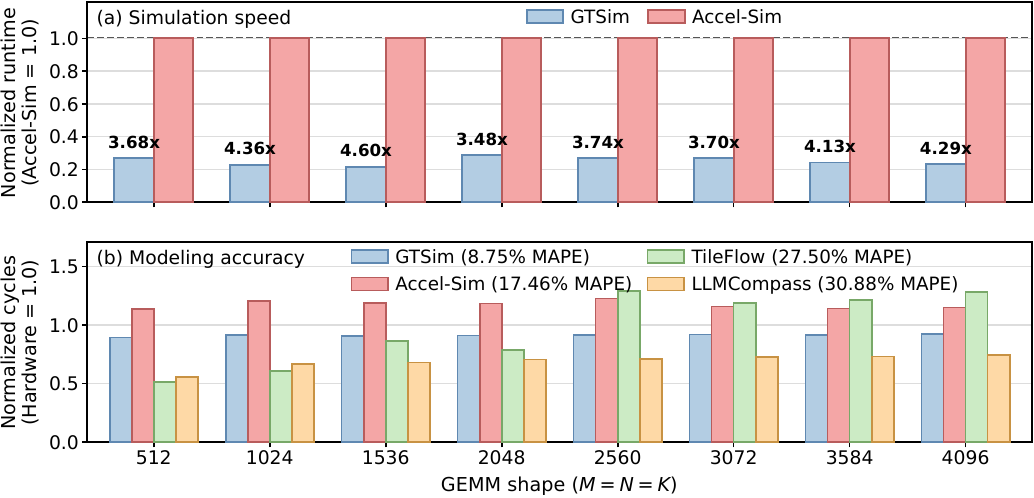}
    \vspace*{-0.4cm}
    \caption{\textcolor{revisionblue}{\Rev{I} Runtime and accuracy comparison on eight A100 square GEMMs. (a) Runtime normalized to Accel-Sim. (b) Simulated cycles normalized to hardware measurements.}}
    \label{fig:runtime-validation}
\end{figure}

\begin{revision}
\Rev{I} \textbf{Simulation efficiency.}
As shown in \Fig{fig:runtime-validation}, we compare eight A100 square GEMMs ($M=N=K=512$--4096) against Accel-Sim, TileFlow, and LLMCompass. \GTSim is 3.48--4.60$\times$ faster than Accel-Sim (3.98$\times$ geometric mean) while maintaining comparable MAPE. TileFlow and LLMCompass are faster because of their coarser abstractions, but their MAPEs increase to 27.50\% and 30.88\%, respectively. These results show that, for representative LLM kernels, tile-level simulation can achieve comparable accuracy to instruction-driven simulation while reducing simulation time.
\end{revision}

\begin{revision}
\Rev{B1} \textbf{Graph construction.}
Graph-construction overhead is kept small for regular kernels through template sharing. Since their CTAs execute isomorphic tile graphs, \GTSim constructs the shared CTA graph structure once rather than replicating it across all logical CTAs. Each logical CTA maintains lightweight metadata and runtime state, including its tile coordinates and execution status, while reusing the same graph structure. Across the evaluated GEMMs, graph construction takes 0.011--0.084\,s and stores 11.52K--88.58K physical nodes representing 0.09M--45.35M logical nodes, with a graph memory footprint of 12.05--91.29\,MiB.
\end{revision}

\begin{table}[t]
\color{revisionblue}
\centering
\footnotesize
\setlength{\tabcolsep}{4pt}
\caption{\textcolor{revisionblue}{\Rev{B2}\ \Rev{E1} Additional H100 validation.}}
\vspace*{-0.2cm}
\label{tab:additional-analysis}
\begin{tabular}{l l r}
\toprule
\textbf{Study} & \textbf{Configuration} & \textbf{MAPE} \\
\midrule
Tile-graph & Full model & 1.55\% \\
ablation & No general order constraints & 5.90\% \\
 & No cross-warp/group sync. & 35.44\% \\
 & Data dependencies only & 42.19\% \\
\midrule
Dynamic & Balanced routing & 13.52\% \\
MoE & 50\% assignments to hottest expert & 8.09\% \\
\bottomrule
\end{tabular}
\vspace*{-0.2cm}
\end{table}

\begin{revision}
\Rev{B2} \textbf{Tile-graph mechanism ablation.} The H100 FP8 GEMM ablation in \Tab{tab:additional-analysis} shows that both general order constraints and cross-warp/group synchronization are necessary for accurate simulation. Removing order constraints allows TMA and epilogue operations without direct data dependence to overlap excessively. Removing synchronization further eliminates consumer barriers, producer--consumer gates, and ping-pong buffer-reuse constraints in each iteration, hiding recurring synchronization costs. Retaining only data dependencies removes both constraint classes and permits the most unrealistic parallelism.
\end{revision}

\begin{revision}
\Rev{E1} \textbf{Dynamic kernel evaluation.} \GTSim supports input-dependent dynamic kernels by accepting runtime workload metadata before graph construction. The frontend uses this metadata to instantiate concrete operation dimensions, tile counts, and node instances; the resulting graph then executes with the same dependency and hardware semantics as a static kernel, without a dynamic-kernel-specific timing path in the backend. We use MoE grouped GEMM as a representative case. For each request or batch, routing determines the token count assigned to each expert and thus the dynamic $M$ dimension of its GEMMs. These per-expert token counts can be collected from a real routing execution and reused across target GPUs because they depend on the model and input rather than the target hardware. We compare with H100 measurements using the Mixtral 8$\times$7B configuration (eight experts, top-2 routing, and 128--4096 input tokens). For $T$ input tokens, top-2 routing produces $2T$ token--expert assignments. Balanced routing divides them equally among the eight experts, whereas skewed routing assigns 50\% to the hottest expert and distributes the remainder as evenly as possible among the other seven. Both FFN up- and down-projection stages execute all experts within one grouped GEMM. As shown in \Tab{tab:additional-analysis}, \GTSim achieves comparable accuracy under both balanced and skewed routing.
\end{revision}

\section{Case Study}
\label{sec:case-study}

This section presents three case studies on software pipelining, NoC-enabled fusion and mapping, and Blackwell adaptation.

\subsection{Software Pipelining Impact}

\textbf{Setup.}
We study a fused \textbf{GEMM + SiLU} persistent kernel, where one thread block processes multiple GEMM tiles wave by wave and the epilogue of one wave can overlap with the mainloop of the next. 
We compare the four different pipeline organizations explained \Sec{sec:tile-graph-frontend} (\Fig{fig:software-pipeline-setup}): (1) \textbf{naive 2-stage}, (2) \textbf{WS cooperative}, (3) \textbf{WS ping-pong}, and (4) \textbf{WS 3-stage}. These organizations differ in how TMA, MMA, and epilogue work are assigned across warp groups and therefore in how much overlap they expose.
They have the same tile configuration $(B_M, B_N, B_K)$ and $K=1024$.
We run each kernel 10 waves, one output tile per wave on a single-SM Hopper configuration. 
WS ping-pong splits each tile into two sub-tiles to overlap the mainloop with the epilogue, increasing TMA traffic in exchange for more overlap. WS 3-stage further places the epilogue on a dedicated warp group; although Hopper does not directly support this behavior, we model it by adding a direct dependency from MMA to the epilogue warp group in the tile graph.

\begin{figure}[t]
    \centering
    \includegraphics[width=1.0\linewidth]{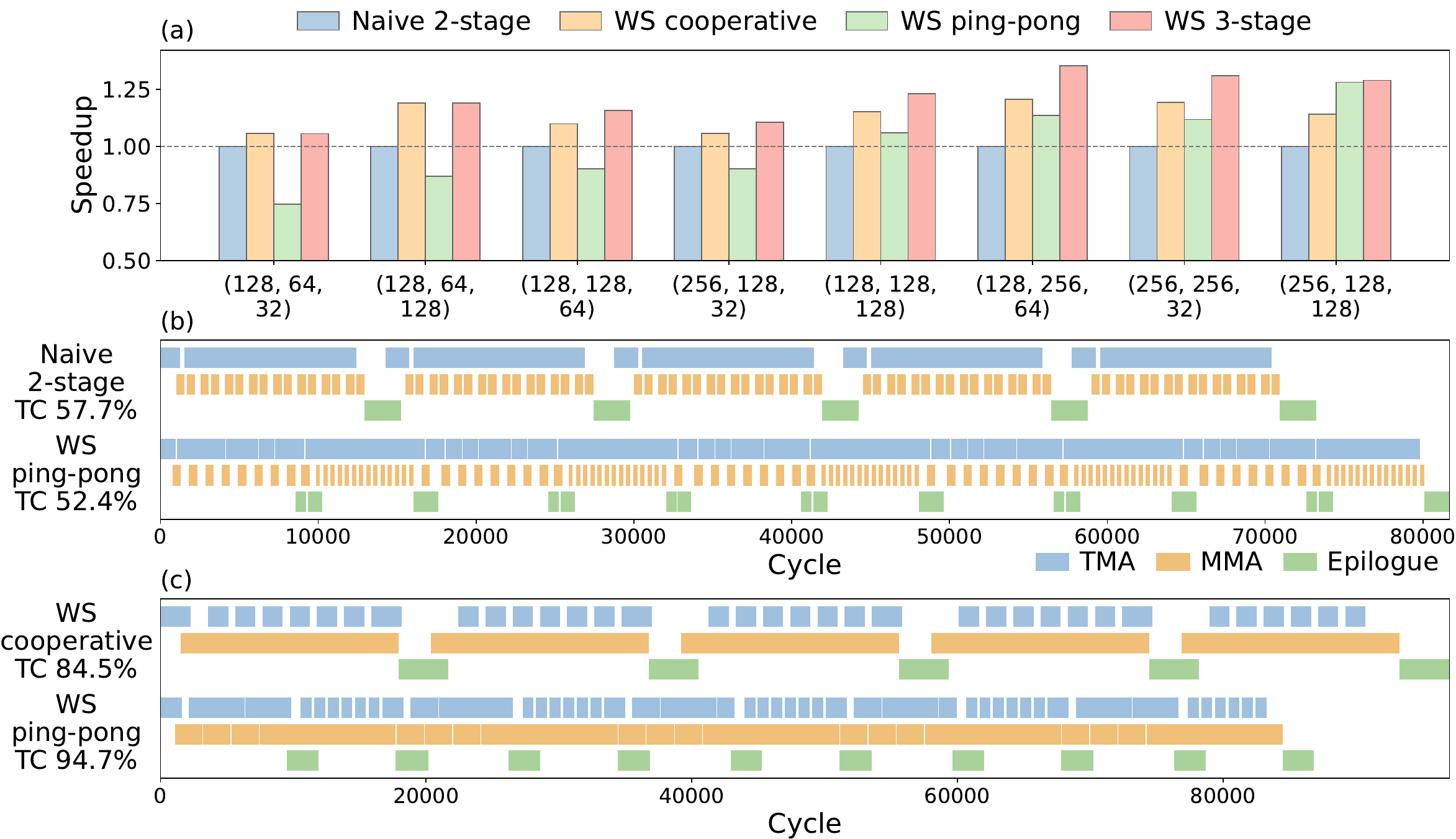}
    \vspace*{-0.2cm}
    \caption{Software pipelining case study. (a) Normalized performance across tile sizes. (b,c) Representative timelines in memory-bound and compute-bound configurations.}
    \label{fig:software-pipeline-performance}
\end{figure}

\textbf{Overall Performance.}
\Fig{fig:software-pipeline-performance}(a) shows normalized performance across different tile sizes. WS cooperative consistently improves over naive 2-stage. WS ping-pong performs worse for small tiles but becomes advantageous for larger tiles. WS 3-stage remains near-optimal across all configurations.

\textbf{Memory-bound vs. Compute-bound Regimes.}
The benefit of pipelining depends on the performance regime. For small tiles, the kernel is memory-bound, as shown in \Fig{fig:software-pipeline-performance}(b) for $(128, 128, 64)$, where Tensor Core utilization is 57.7\% for naive 2-stage and 52.4\% for WS ping-pong. In this regime, WS ping-pong underperforms because tile splitting increases TMA traffic, and the extra memory cost outweighs the additional overlap. WS cooperative still improves over naive 2-stage by separating TMA and MMA across warp groups and reducing stalls.

For larger tiles, the kernel becomes compute-bound, as shown in \Fig{fig:software-pipeline-performance}(c) for $(256, 256, 128)$, where Tensor Core utilization is 84.5\% for WS cooperative and 94.7\% for WS ping-pong. Here the longer MMA phase hides TMA latency effectively, so the overlap benefit dominates and WS ping-pong outperforms WS cooperative.

\textbf{WS 3-stage Pipeline.}
WS 3-stage remains near-optimal by fully decoupling epilogue from the mainloop using a dedicated warp group. This enables complete overlap without introducing additional TMA traffic, while also allowing larger effective tile sizes and high Tensor Core utilization, reaching 65.0\% in the memory-bound case and 93.7\% in the compute-bound case.

\textbf{Takeaway.}
These results show that the effectiveness of software pipelining depends on the performance regime. In memory-bound scenarios, minimizing memory overhead is critical, while in compute-bound scenarios, maximizing overlap becomes more important. Different pipeline designs correspond to different dependency structures in the tile graph, enabling \GTSim to capture their performance impact.

\subsection{NoC-enabled Fusion and Mapping}

% 利用noc的更好的数据流？ wafer llm

\textbf{Motivation. }
Modern GPUs provide on-chip communication (e.g., DSM/NoC) that enables direct data exchange across thread blocks. This allows intermediate data to remain on-chip instead of being written to DRAM. In addition, NoC topology and mapping can significantly impact communication efficiency.

\textbf{Setup. }
We consider a fused attention pipeline including RMSNorm, QKV projection, RoPE, FlashAttention, rescale, and output projection, following a cluster-based execution model similar to ClusterFusion~\cite{luo2025clusterfusion}. Each attention head is mapped to a cluster, where thread blocks exchange intermediate data via NoC using remote load/store primitives. Communication is modeled with four \textit{cluster reduce} operations and one \textit{cluster gather}: one reduce after RMSNorm, one gather after QKV projection, two reduces inside FlashAttention, and one reduce after rescale. The default NoC is a 16-node Hopper-like crossbar; we also model a 4$\times$4 mesh with the same average latency and total bandwidth.

\begin{figure}[t]
    \centering
    \includegraphics[width=1.0\linewidth]{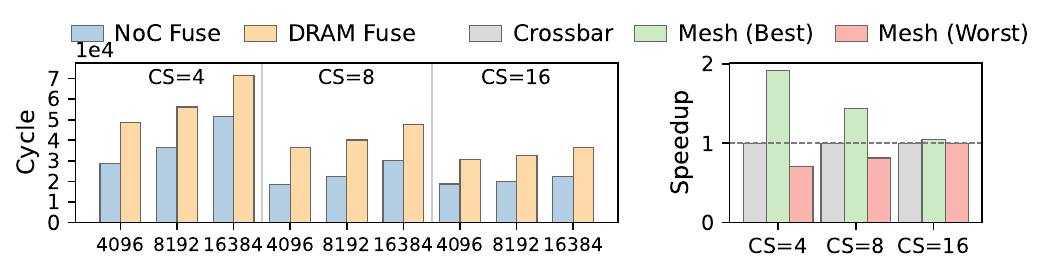}
    \vspace*{-0.2cm}
    \caption{Left: total kernel cycles for NoC-based and DRAM-based communication. Right: normalized NoC residence time under different topologies and mappings.}
    \label{fig:noc-result}
\end{figure}

\textbf{Results.}
\Fig{fig:noc-result} shows the impact of NoC-enabled kernel fusion and topology-aware mapping with two following insights.

\textbf{(1) NoC vs. DRAM.}
\Fig{fig:noc-result}-left shows that using NoC for intermediate data exchange consistently reduces execution time compared to DRAM-based communication, with an average speedup of $1.65\times$ and a maximum of $1.94\times$. 
%The gain comes from keeping collective communication on-chip and avoiding repeated DRAM accesses.

\textbf{(2) Topology-aware Mapping.}
\Fig{fig:noc-result}-right shows the impact of NoC topology and mapping through the \textit{average NoC residence time per request}. While the crossbar provides uniform latency, the mesh exhibits significant variation depending on placement. On average, topology-aware mapping reduces NoC residence time by $1.47\times$, with the best case reaching $1.91\times$. The benefit is largest at small cluster sizes, where placement can avoid long paths; as the cluster grows and occupies the full mesh, mapping flexibility decreases and performance converges toward the topology average.

\textbf{Takeaway.}
This case study shows that \GTSim can capture the performance impact of inter-SM communication design and topology-aware mapping. In particular, it models not only the benefit of NoC-enabled fusion, but also how communication topology and block placement jointly affect fused-kernel efficiency.

\subsection{Adapting to Latest Blackwell Architecture}

\textbf{Adapting to Blackwell.}
\begin{revision}
\Rev{III} We adapt our simulator from Hopper to Blackwell (B200) by extending the hardware model, operation set, and graph semantics. Specifically, we add TMEM as a resource model, introduce \texttt{tcgen05} and TMEM-access operations, and add grouped/CTA-pair issue semantics for Blackwell Tensor Core execution. Relative to the immediate pre-Blackwell simulator core, this extension modifies 11.3\% of the simulator core. The changes are localized to operation and graph metadata, the TMEM model, grouped issue handling, SM/memory integration, and architecture configuration. The generic dependency-release mechanism, threadblock residency framework, ready-node scheduling framework, compute-pipeline abstraction, TMA and NoC models, and existing L2/DRAM models are reused.
\end{revision}

\begin{figure}[t]
    \centering
    \includegraphics[width=0.95\linewidth]{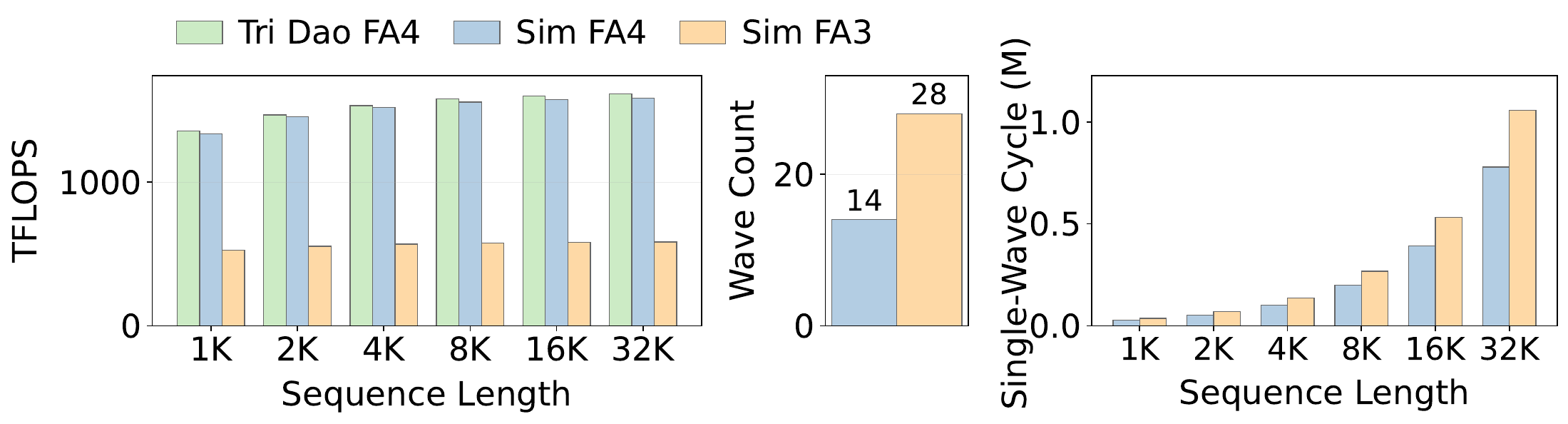}
    \vspace*{-0.2cm}
    \caption{FA3 vs.\ FA4 on B200. Left: TFLOPS. Middle: wave count. Right: normalized single-wave cycles.}
    \label{fig:fa4-performance}
\end{figure}

\textbf{Setup.}
\begin{revision}
\Rev{C2} Because local B200 measurements are currently unavailable, we evaluate FlashAttention-4 (FA4) against performance reported in the published FA4 paper~\cite{flashattention4} and compare it with FA3~\cite{flashattention3} under the same B200 configuration. The total token count is fixed to 32K and the head dimension is 128. We implement a simplified FA4 based on the published design. To enable a fair comparison on Blackwell, we adapt FA3 to the B200 architecture by replacing its compute and memory primitives with Blackwell-compatible operations while keeping the original tile sizes unchanged. \Fig{fig:fa4-performance} (left) shows comparable performance to the published FA4 values, providing only a coarse, preliminary validation on Blackwell.
\end{revision}

\textbf{FA3 vs. FA4 on Blackwell.}
\Fig{fig:fa4-performance} shows that FA4 significantly outperforms FA3 across all sequence lengths. Using \GTSim, we derive two following reasons to explain FA4's performance advantage on Blackwell over FA3 that is originally designed for Hopper.
%This gap can be explained by two compounding factors.

\textbf{(1) Inter-threadblock granularity (wave count).}
As \Fig{fig:fa4-performance} (middle) shows, FA4 requires only 14 waves, while FA3 requires 28 waves for the same workload. This $2\times$ difference comes from FA4's larger per-threadblock tile size. On Blackwell with a greater TC throughput, larger tiles are needed to keep the compute pipeline fully utilized; smaller tiles lead to underutilization and higher scheduling overhead.

\textbf{(2) Intra-threadblock efficiency (single-wave execution).}
\Fig{fig:fa4-performance} (right) shows that FA4 also achieves about $1.3\times$ lower single-wave cycle count. This improvement comes from Blackwell's TMEM together with FA4's explicit three-way warp-group specialization, and is further aided by FA4's SIMD polynomial approximation for exp, which helps distribute the exp bottleneck across regular arithmetic units. In FA4, MMA results can be materialized through TMEM and consumed by a separate warp group, allowing MMA, softmax/update, and memory movement to proceed in a more decoupled manner. In FA3, softmax and MMA remain more tightly coupled, interrupting tensor issue and reducing tensor-pipeline continuity.

\textbf{Takeaway.}
This case study demonstrates that our simulator can be easily extended to new architectures while preserving predictive accuracy. More importantly, it enables analysis of how architectural changes interact with kernel design. In particular, Blackwell favors larger execution granularity and deeper pipeline decoupling, highlighting the importance of co-designing kernels with emerging hardware features.

\section{Related Work}

\textbf{Instruction-driven simulators } are standard approaches for architecture exploration. Gem5~\cite{binkert2011gem5} is widely used for CPU simulation and gem5-gpu~\cite{power2015gem5gpu} extends it to heterogeneous CPU-GPU systems. In the accelerator domain, SCALE-Sim~\cite{samajdar2020scalesim}, mNPUsim~\cite{hwang2023mnpusim}, and PyTorchSim~\cite{yang2025pytorchsim} target systolic-array accelerators, multi-core NPUs, and broader NPU execution scenarios, respectively. In the GPU domain, GPGPU-Sim~\cite{bakhoda2009gpgpusim}, Accel-Sim~\cite{khairy2020accelsim}, MGPUSim~\cite{sun2019mgpusim}, NVArchSim~\cite{villa2021nvarchsim}, and Vortex~\cite{tine2021vortex} respectively target NVIDIA GPU, its newer generation, multi-GPU systems, faster system-level simulation, and RISC-V-based GPU architecture. These simulators enable detailed evaluation across diverse architectures, but they are typically tied to architecture-specific low-level semantics and hardware structures, making adaptation to new hardware generations costly.

\textbf{Analytical models } provide a faster alternative for performance estimation and design-space exploration. 
GPUMech~\cite{huang2014gpumech} first introduces interval analysis into GPU performance modeling, and MDM~\cite{wang2020mdm} emphasizes memory divergence.
GCoM~\cite{cha2022gcom} incorporates more detailed modern GPU core structure and workload imbalance, and AMALI~\cite{cao2025amali} extends this direction to LLM inference with improved TC and cache modeling. 
Interval-based analytical models are useful for bottleneck analysis, but still rely on instruction-level traces and representative-warp style abstractions, making them difficult to model kernels with heterogeneous warp behavior such as warp specialization. 
In mapping-based modeling, Timeloop~\cite{parashar2019timeloop}, MAESTRO~\cite{kwon2019maestro}, TENET~\cite{lu2021tenet}, TileFlow~\cite{zheng2023tileflowfusion}, and LLMCompass~\cite{zhang2024llmcompass} respectively emphasize mapping and loop-nest analysis, data-centric reuse modeling, improved latency estimation, tile-centric fusion modeling, and LLM-oriented mapping and scheduling search with a hardware cost model. These models are well suited to design-space exploration, but are generally too coarse to explicitly capture fine-grained asynchronous execution, inter-warp coordination, and new GPU hardware features.

\section{Conclusion}

In this paper, we present \GTSim, a tile-centric GPU simulation framework that models kernel execution as a dependency-driven warp-centric tile graph and combines an automatic frontend with a graph-driven backend using throughput-oriented compute, memory, and NoC models. \GTSim can accurately model representative GEMM, attention, and end-to-end LLM inference workloads, capture the performance impact of software pipelining and NoC-enabled fusion, and be extended to emerging architectures such as Blackwell with localized changes. We plan to open-source it to facilitate future LLM hardware-software co-design.

%Overall, these results show that \GTSim is a practical tool for GPU hardware-software co-design.
%Modern GPU kernels for LLM workloads increasingly rely on explicit dependencies among tile-level computation, data movement, and synchronization, while recent architectures continue to evolve rapidly. 

% \input{src/10-acknowledge}

%%%%%%% -- PAPER CONTENT ENDS -- %%%%%%%%

\bibliographystyle{ACM-Reference-Format}
\bibliography{reference}

\end{document}